\documentclass[fleqn,twoside]{article}

\usepackage{espcrc2,isolatin1}

\usepackage{graphicx}
\usepackage{amsmath}

\def\eq#1{Eq.\ (\ref{#1})}

\def\sec#1{Section \ref{#1}}

\def\fig#1{Fig.~\ref{#1}}

\newcommand{\lsim}{ {\
\lower-1.2pt\vbox{\hbox{\rlap{$<$}\lower5pt\vbox{\hbox{$\sim$}}}}\ } }
\newcommand{\gsim}{ {\
\lower-1.2pt\vbox{\hbox{\rlap{$>$}\lower5pt\vbox{\hbox{$\sim$}}}}\ } }

\def\l{\left}
\def\r{\right}
\def\vec#1{\mathbf{#1}}

\newcommand{\be}{\begin{equation}}
\newcommand{\ee}{\end{equation}}
\newcommand{\bea}{\begin{eqnarray}}
\newcommand{\eea}{\end{eqnarray}}

\newcommand{\nn}{\nonumber}
\newcommand{\ra}{\rightarrow}

\newcommand{\cZ}{\mathcal{Z}}

\newcommand{\cR}{\mathcal{R}}

\newcommand{\la}{\langle}
\renewcommand{\ra}{\rangle}

\newcommand{\Tr}{\mbox{\rm Tr}}

\newcommand{\mev}{\mbox{\rm MeV}}
\newcommand{\gev}{\mbox{\rm GeV}}

\def\ods2{\mathcal{O}_{\Delta S=2}}
\def\zds2{Z_{\Delta S=2}}

\title{
{\vspace{-3.45cm} \normalsize \hfill
\parbox{35mm}{\hfill CPT-2003/P.4582\\ \phantom{x}\hfill BU-HET 03-20 \\ \phantom{x}\hfill 
BNL-HET-03/21}
}\\[22mm]
Preliminary results from a simulation of quenched QCD with overlap
fermions on a large lattice\thanks{Work supported in part by US DOE
grants DE-FG02-91ER40676 and DE-AC02-98CH10866, EU HPP contracts
HPRN-CT-2000-00145 and HPRN-CT-2002-00311, and grant
HPMF-CT-2001-01468.  We thank Boston University and NCSA for use of
their supercomputer facilities.}\thanks{Combined presentations by 
C.~Hoelbling, L.~Lellouch and C.~Rebbi at {\it Lattice 2003}, Tsukuba, Japan. \newline
$^\ddag$ Unit\'e Propre de Recherche 7061.}
}

\author{
F. Berruto\address[bnl]{Department of Physics, Brookhaven National
Laboratory, Upton NY 11973, USA
\vspace{-0.2cm}},
N. Garron\address[cpt]{Centre de Physique
Th\'eorique$^\ddag$, Case 907, CNRS Luminy, F-13288 Marseille Cedex 9, France
\vspace{-0.2cm}},
C. Hoelbling\addressmark[cpt],
L. Lellouch\addressmark[cpt],
C. Rebbi\address[bu]{Department of Physics, Boston University, 590
Commonwealth Avenue, Boston MA 02215, USA
\vspace{-0.2cm}}
and N. Shoresh\addressmark[bu]
}

\begin{document}

\begin{abstract}
We simulate quenched QCD with the overlap Dirac operator.  We work
with the Wilson gauge action at $\beta=6$ on an 
$18^3\times64$ lattice. We calculate quark propagators for a single source
point and quark mass ranging from $am_q=0.03$ to $0.75$.
We present here preliminary results based on the propagators
for 60 gauge field configurations.
\vspace{-0.5cm}
\end{abstract}

\maketitle

\section{Introduction}

\vspace{-0.2cm}

The closely related domain wall~\cite{Kaplan:1992bt,Shamir:1993zy}
and overlap~\cite{Narayanan:1995gw,Narayanan:1994sk,Neuberger:1998fp}
formulations of lattice fermions, with their ability to preserve chiral
symmetry even at finite lattice spacing~\cite{Ginsparg:1982bj,Luscher:1998pq},
offer an almost ideal tool for lattice QCD calculations.
Still overlap or domain wall fermions can only be implemented
at a high computational cost.  Thus it is important to perform
exploratory calculations on large lattices in order to validate
the applicability of these novel formulations.  At the same time,
large scale QCD simulations with overlap or domain wall fermions,
because of the benefits of chiral symmetry, can produce valuable
cross-checks of observables calculated with more traditional
quark discretizations, or even permit the evaluation of observables
otherwise out of the reach of practical calculations.  In earlier
studies we simulated quenched QCD with overlap fermions on a
$16^3\times32$ lattice, obtaining results for the pseudoscalar
spectrum, strange quark mass and quark
condensate~\cite{Giusti:2001pk,Giusti:2001yw},
for the kaon $B$ parameter~\cite{Garron:2002dt,Garron:2003cb} and
for non-perturbative
renormalization constants~\cite{Giusti:2001yw,Garron:2002dt,Garron:2003cb}.
Although the results of this earlier work turned out to be quite
satisfactory, the size of the lattice, especially its extent in time,
were found insufficient for the calculation of some observables.
For example, the plateaus needed to isolate the $K$ mesons in the
calculation of $B_K$ extended for only two lattice spacings and
we could not reliably evaluate vector meson masses and decay constants
or baryon masses.  This prompted us to extend our overlap calculations to
a larger system.  It is important to observe that increasing the
size of the lattice in simulations with overlap fermions does not only
entail an augmentation of the (already quite heavy) computational cost.
Indeed, since the calculation of the overlap operator requires the
use of some suitable approximation to the inverse square root of a very
large sparse matrix, the increase of the size of the matrix may produce
serious problems of convergence.  For these reasons, and also in
consideration of the computational resources available to us,
we decided to use a lattice of size $18^3\times 64$, with double
the extent in time, but only a moderate increase in the spatial extent
with respect to our former calculation.  Thus we generated
and archived 100 $18^3\times 64$ pure gauge field configurations
with the Wilson gauge action at $\beta=6$.  We used a 6-hit Metropolis
algorithm, tuning the acceptance to $\approx 0.5$, performed 11,000
initial equilibrating iterations, after which 10,000 iterations were
done between each pair of subsequent configurations.  From
a measurement of Wilson loops we determined $r_0/a=5.36\pm0.11$
for the Sommer scale defined by $r_0^2 F(r_0)=1.65$.  With
$r_0=0.5{\rm fm}$ we get $a^{-1}=2.11\pm0.04{\rm GeV}$.

For each of the above configurations we calculated quark propagators
with the overlap Dirac operator, with $\rho=1.4$, for a single point
source and all 12 color-spin combinations, for $am_q=0.03,
0.04, 0.06, 0.08. 0.1, 0.25, 0.5, 0.75$.  For brevity we do not
reproduce here the formula for the overlap Dirac operator: we
refer to~\cite{Giusti:2001pk,Giusti:2001yw} for the relevant equations
and selected citations.  The results we present here are based
on the analysis of 60 configurations.  For the calculation of
the overlap operator (more properly, of its action on a given vector)
for the first 55 configurations we used the Zolotarev approximation
with 12 poles, after Ritz projection of the lowest 12 eigenvectors
of $H^2$.  We found, however, the Chebyshev approximation to be
computationally less costly (by approximately 20\%) and therefore
we are now using this approximation.  With both approximations
we impose $\vert D^\dag D \psi -\chi \vert^2 < 10^{-7}$ as convergence
criterion.  We project out the lowest 12 eigenvectors of $H^2$
also with the Chebyshev approximation and use the evaluation
of the next highest eigenvalue to set the required degree of the expansion,
which typically varies between 100 and 500.  Our code has been written
in F90 with OMP directives and we have been running in shared memory
mode on 16 and 32 processor IBM-P690 nodes at Boston University and
NCSA.

 To conclude this section, we would like to mention that other large
scale quenched QCD calculations with overlap fermions have been presented
in~\cite{Dong:2003zf,Dong:2003im,Chiu:2003iw,DeGrand:2003in}.

\vspace{-0.2cm}

\section{Light hadron spectrum and quark condensate}
\label{sect:lhs}
\vspace{-0.2cm}
Our first concern has been to verify that the results with the
new, larger lattice are compatible with the values found for the
observables in our former calculations~\cite{Giusti:2001pk,Giusti:2001yw}
and that the measurement of the observables can be extended to the
much larger temporal extent with reliable statistics.  We present
here a sample of these checks.

\begin{figure}[!ht]
\vspace{-0.0cm}
\includegraphics*[width=7.5cm]{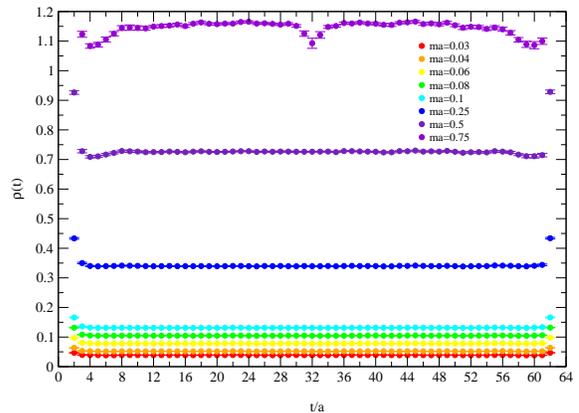}%
\vspace{-1.0cm}
\caption{Results for the axial Ward identity. \label{fig:rhot}}
\vspace{-0.6cm}
\end{figure}

In Fig.~\ref{fig:rhot} we report the new results for the ratio
\be
\rho(t)=G_{\nabla_0 A_0 P}(t)/G_{P P}(t)
\label{rhot}
\ee
where $G_{IJ}(t)$, $IJ=\nabla_0A_0P, PP$, are non-singlet zero-momentum, 
2-point functions
made from the bilinears $I$ and $J$.
On account of the axial Ward identity, $\rho(t)$ should be constant in time.
The figure shows that $\rho(t)$ can be reliably measured over the
entire extent of the lattice for all quark mass values and is indeed
constant in time.
\begin{figure}[!ht]
\vspace{-0.4cm}
\includegraphics*[width=7.5cm]{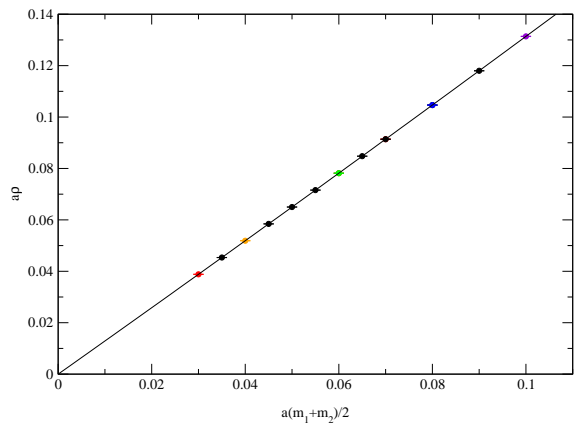}%
\vspace{-1.0cm}
\caption{The ratio $\rho$ of Eq.~\ref{rhot} as function
of quark mass. \label{fig:rho}}
\vspace{-0.6cm}
\end{figure}
The fit $\rho=A+a(m_1+m_2)/Z_A+C (am_1+am_2)^2$ (see Fig.~\ref{fig:rho})
gives $A=-(0.96\pm3.35)\times 10^{-5}, Z_A=1.554\pm0.001\;[1.55\pm0.04],
C=(6.87\pm0.18) \times 10^{-2}$.  The numbers in square brackets, here and
in the following, indicate results we found
in~\cite{Giusti:2001pk,Giusti:2001yw}.

\begin{figure}[!ht]
\vspace{-0.4cm}
\includegraphics*[width=7.5cm]{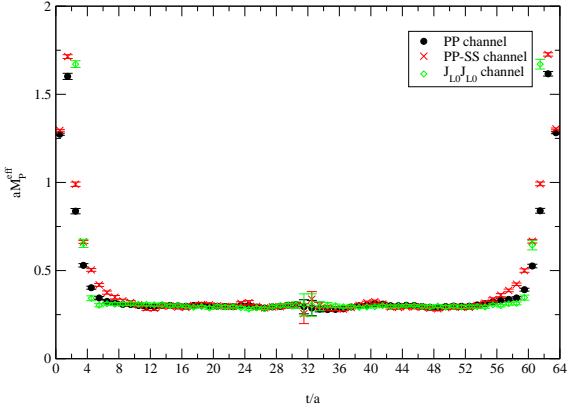}%
\vspace{-1.0cm}
\caption{Effective pseudoscalar mass plateaus for $am_q=0.06$.
\label{fig:meffppss}}
\vspace{-0.6cm}
\end{figure}

Figure~\ref{fig:meffppss} illustrates the value of the effective
mass from the (0-momentum component of the) correlation functions of
the pseudoscalar density, of the pseudoscalar minus scalar densities
and of the temporal component of the left-handed currents, for $am_q=0.06$.
We see that the signal for the correlation functions extends through
the whole lattice.
\begin{figure}[!ht]
\vspace{-0.4cm}
\includegraphics*[width=7.5cm]{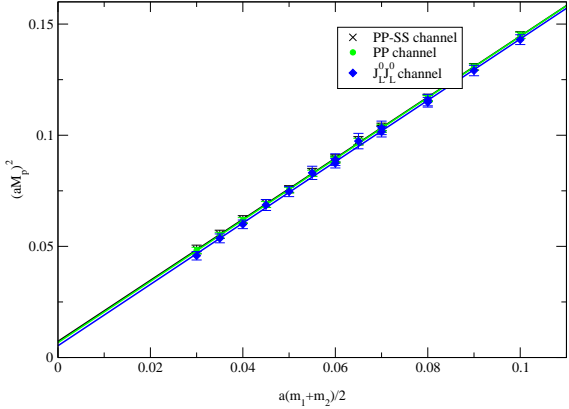}%
\vspace{-1.0cm}
\caption{Pseudoscalar mass squared as function of the total quark mass.
\label{fig:mqmp2}}
\vspace{-0.9cm}
\end{figure}
These three correlation functions, symmetrized with respect to
$t\to 64a -t$, have been fit in the range $12\le t/a \le 32$, and
Fig.~\ref{fig:mqmp2} displays the values thus found for the pseudoscalar
mass as well as the linear fit $(aM_P)^2=A+B a(m_1+m_2)/2$ which gives
\bea
{\rm PP:} \;A=(6.69\pm1.76) \times 10^{-3},\, B=1.38\pm0.02 \nonumber \\
{\rm PP-SS:} \;A=(7.27\pm2.44) \times 10^{-3},\, B=1.37\pm0.02 \nonumber \\
{\rm J_L^0, J_L^0:} \;A=(5.31\pm2.58) \times 10^{-3},\, B=1.38\pm0.04
\nonumber \\
{[{\rm PP:}\; B=1.39\pm0.03;\;\;{\rm PP-SS:}\; B=1.43\pm0.07]}  \nonumber
\eea

\begin{figure}[!ht]
\vspace{-0.0cm}
\includegraphics*[width=7.5cm]{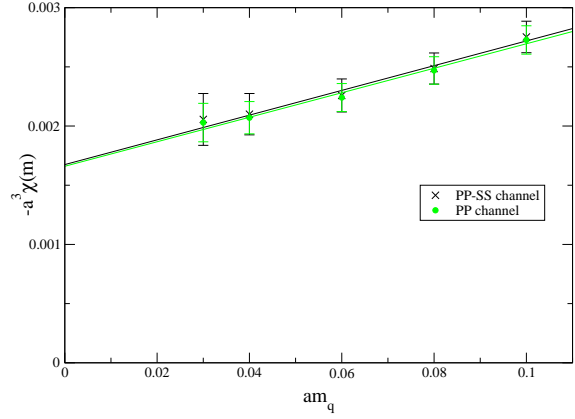}%
\vspace{-1.0cm}
\caption{Determination of the chiral condensate.\label{fig:gmor2}}
\vspace{-0.6cm}
\end{figure}
Figure~\ref{fig:gmor2} shows the results for the scalar
condensate from the Gell-Mann, Oakes and Renner relation, for degenerate
quark masses, together with the linear fit $\chi=A+B am_q$, which gives
\bea
{\rm PP:} \;A&=&(1.66\pm0.18) \times 10^{-3} \nonumber \\
 B&=&(1.03\pm0.15) \times 10^{-2} \nonumber \\
{\rm PP-SS:} \;A&=&(1.67\pm0.24) \times 10^{-3} \nonumber \\
B&=&(1.04\pm0.21) \times 10^{-2} \nonumber \\
{[}{\rm PP-SS:} \;A&=&(1.17\pm0.27) \times 10^{-3} {]}\nonumber
\eea

\begin{figure}[!ht]
\vspace{-0.4cm}
\includegraphics*[width=7.5cm]{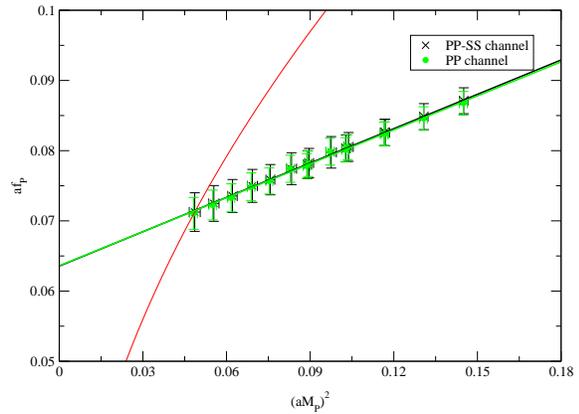}%
\vspace{-1.0cm}
\caption{Determination of $a$. \label{fig:ainv}}
\vspace{-0.6cm}
\end{figure}
Finally, we reproduce in Fig.~\ref{fig:ainv} the determination of $a$
with the method of lattice physical planes: the red line
corresponds to the physical ratio $F_K/M_K$ with $M_K=0.495\,\gev$
$F_K=0.16\,\gev$.  We find $a^{-1}=2.238 \pm0.0 83\,\gev$ from
the PP channel and  $a^{-1}=2.236\pm0.098 [2.29\pm0.09]\,\gev$ from
the PP-SS channel.

\begin{figure}[!ht]
\vspace{-0.4cm}
\includegraphics*[width=7.5cm]{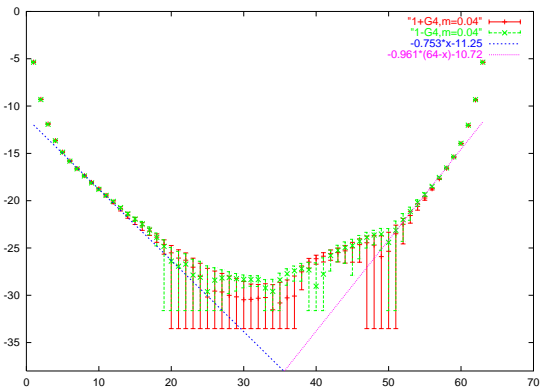}%
\vspace{-1.0cm}
\caption{Logarithm of the baryon-baryon correlation functions.
\label{fig:N004}}
\vspace{-0.6cm}
\end{figure}

\begin{figure}[!ht]
\vspace{-0.65cm}
\includegraphics*[width=7.5cm]{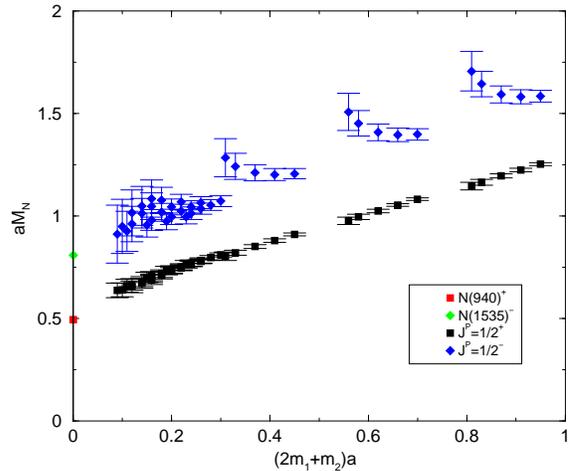}%
\vspace{-1.0cm}
\caption{Baryon masses as function of the total quark mass. \label{fig:NpNn}}
\vspace{-0.6cm}
\end{figure}

\smallskip
Figures~\ref{fig:N004},~\ref{fig:NpNn} illustrate new results, which
were out of reach with a smaller lattice.  In Fig.~\ref{fig:N004} we
show the quantity $\log|G_{BB}(t)|$ with
$B\equiv B_\pm=(1\pm\gamma_0) \epsilon_{abc}
\left[u_a^T C\gamma_5 d_b \right]u_c$ (a sum over the spatial coordinates
of $B$ over the time slice at $t$ is implicit).  Since the
correlation function with the $1+\gamma_0$ projector describes
forward propagating $1/2^+$ baryons and backward propagating $1/2^-$
baryons, while the two parities are interchanged with the $1-\gamma_0$
projector, in Fig.~\ref{fig:N004} we plot the $1-\gamma_0$
correlation function against $64a - t$.  The results for the two
correlation functions, after this inversion of the time axis,
are quite consistent, as one would expect, and exhibit an extended linear
region over which one can extract an estimate of the lowest baryon
masses.  The corresponding linear fits are also shown in Fig.~\ref{fig:N004}.

In Fig.~\ref{fig:NpNn} we collect all of our results for the lowest
$1/2^+$ and $1/2^-$ baryon masses.  We calculated the masses
corresponding to all combinations where two of the quarks have the
same mass.  The masses of the $1/2^+$ states appear to fall on a
single curve, indicating a dependence only on the total quark mass.
The masses of the $1/2^-$ seem to indicate a dependence on the quark
mass difference also, but the statistical errors are high and a more
elaborated analysis should be done before deriving any such
implication from the data.  We emphasize that these results are
preliminary.  In particular the masses in Fig.~\ref{fig:NpNn} have
been obtained with point sources {\it and} sinks.  While having
calculated point source quark propagators forces us to use point
sources, we should be able to improve the statistical significance of
the results by using suitable extended sink operators.  Calculations
along these lines are in progress.

\vspace{-0.2cm}

\section{Experimenting with the charm quark}

\vspace{-0.2cm}

In the present section we investigate the behavior of overlap
fermions with heavy quarks $Q$ for which
$am_Q\hspace{4pt}/\hspace{-10pt}\ll 1$. In particular, we are
interested in understanding whether the charm quark can be
approximately simulated at lattice spacings of around $2\,\gev$, where
the charm has a bare mass $am_c\sim 0.75$. Related results, obtained
with a different gauge action, can be found in \cite{Liu:2002qu}.

Though a rigorous investigation of discretization errors would require
simulations at a number of values of the lattice spacing, at the
present time we only have a simulation for which $a^{-1}\sim
2\,\gev$. We have thus devised a number of tests which provide us with
some measure of the mass-dependent discretization errors present in
our simulation. They concern a variety of quantities, some of which
are of direct phenomenological interest such as the leptonic decays
constant of the $D_s$ meson, $f_{D_s}$.

\vspace{-0.2cm}

\subsection{Axial and vector Ward identities}

We turn once again to the non-singlet axial (AWI) and vector (VWI) Ward
identities. Instead of working with the ratio $\rho(t)$ defined in \eq{rhot},
we consider the quantity
\be
\cZ_A(t;m_1,m_2)\equiv
\frac{(am_1+am_2)}{\rho(t;m_1,m_2)}
\ ,
\label{eq:cZAdef}
\ee
with a similar definition for $\cZ_V(t;m_1,m_2)$ in terms of vector
currents and scalar densities. These quantities are also constant in
time, by virtue of the AWI and VWI, and reduce to the renormalization
constants $Z_A$ and $Z_V$ of the axial and vector currents in the
chiral limit.  $\cZ_A(t)$ and $\cZ_V(t)$ have very nice plateaus in
$t$ for all quark-mass combinations, though error bars and
fluctuations on $\cZ_V(t)$ are substantially larger. We thus fit them
to a constant in the time range $14\le at\le 50$, thereby obtaining
mass-dependent ``renormalization constants'' $Z_A(m_1,m_2)$ and
$Z_V(m_1,m_2)$. The fits to $\cZ_A(t)$ are all excellent, except for
the mass combination $am_1=am_2=0.75$ where there is some evidence
that the signal is lost for a handful of points around the center of
the lattice (cf.\ \fig{fig:rhot}). The situation is much less
satisfactory for $\cZ_V(t)$, where the fluctuations make it difficult
to obtain reliable fits.

\begin{figure}[!ht]
\vspace{-0.0cm}
\includegraphics*[width=7.5cm]{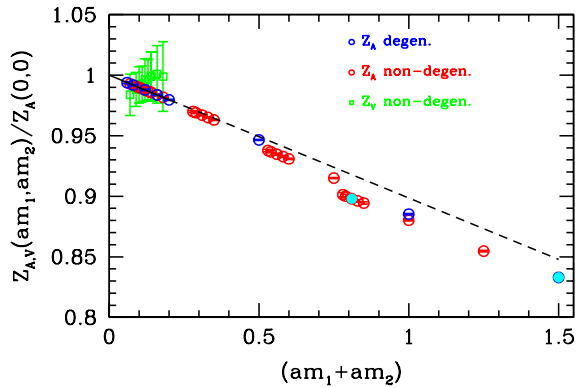}%
\vspace{-1.0cm}
\caption{\label{fig:ZAVvsam1pm2} $Z_{A,V}(m_1,m_2)/Z_A(0,0)$ vs.\ $(am_1+am_2)$.
The straight line corresponds to the fit described
around \eq{eq:ZAmmfit}. The two filled symbols approximately
correspond, from left to right, to a $D_s$ and an $\eta_c$ meson.}
\vspace{-0.6cm}
\end{figure}

The quantities $Z_{A,V}(m_1,m_2)$ are of particular interest here,
because their mass dependence is solely due to discretization errors. Given
our definition, the leading mass dependence is of the form
$a^2(m_1+m_2)$ and $a^2(m_1-m_2)^2/(m_1+m_2)$.  In
\fig{fig:ZAVvsam1pm2} we plot $Z_A(m_1,m_2)/Z_A(0,0)$ as a function of
$(am_1+am_2)$ for all mass combinations, were $Z_A(0,0)$ is obtained
as described below. We consider $Z_V(m_1,m_2)$ only for
$(am_1+am_2)\le 0.2$, since it is not clear that we can reliably fit
$\cZ_V(t)$ for larger masses. The values of $Z_A(m_1,m_2)$ for
$(am_1+am_2)\le 0.2$ display linear behavior, indicating that
mass-dependent discretization errors are dominated by $a^2(m_1+m_2)$-type
terms in this range of masses. Therefore, in this mass range we fit
$Z_A(m_q,m_q)$ to:
\be
Z_A(m_q,m_q)=Z_A(0,0)\l(1+2(aB)(am_q)\r)
\ ,
\label{eq:ZAmmfit}\ee
where we only consider degenerate combinations to eliminate possible
$a^2(m_1-m_2)^2/(m_1+m_2)$ terms. We find $Z_A(0,0)=1.5544(15)$, which
is entirely compatible with the value given after \eq{rhot}, and
$aB=-0.102(5)$.

The first point worth noting is that the coefficient $B$ of the
$a^2(m_1+m_2)$ discretization error is small, $B\sim 225\,\mev$. The
second interesting feature is that deviation from linearity becomes
statistically significant (i.e.\ larger than 3 standard deviations) only
for $(am_1+am_2)\gsim 0.28$.  We are actually able to consistently fit
$Z_A(m_1,m_2)$ all the way up to $am_1=am_2=0.5$ with the addition of
two terms proportional to $a^3(m_1+m_2)^2$ and
$a^2(m_1-m_2)^2/(m_1+m_2)$, and all the way up to $am_1=am_2=0.75$
with an additional term proportional to $a^4(m_1+m_2)^3$, though the
coefficients obtained in the latter fit are only marginally consistent
with those from our two other fits. In these fits to extended mass
ranges, we further find that the coefficients of the higher order
terms are approximately one half to a full order of magnitude
smaller than $aB$. The net result is that discretization errors
inferred from $Z_A(m_1,m_2)$ are approximately 10\% at
$(am_1,am_2)=(0.06,0.75)$ and 15\% at $(am_1,am_2)=(0.75,0.75)$, 
mass combinations which corresponds roughly to a $D_s$ and an $\eta_c$ meson,
respectively.
Moreover, chiral symmetry seems to be good, as evidenced by the
compatibility of $Z_{A}(m_1,m_2)$ and $Z_{V}(m_1,m_2)$ at small
masses. Thus, there is no evidence for a major breakdown of overlap
fermions up to $am_q=0.75$ and $(am_1+am_2)=1.5$.

\vspace{-0.2cm}

\subsection{2-point functions, pseudoscalar meson masses and decay constants}

We have performed fits to the zero-momentum $PP$, $PP{-}SS$, $A_0P$, $PA_0$, $A_0A_0$ and
$J_0^LJ_0^L$ correlators, all of which have the pseudoscalar meson as
the lightest contributing state. The fit form used is:
\bea
&&\sum_{\vec x}\la I(x)J(0)\ra\stackrel{ax_0\gg 1}{\longrightarrow}\label{eq:2ptfndef}\\
&&
\frac{\cZ_{IJ}}{2M_{IJ}}\l[e^{-M_{IJ}x_0}+s_{IJ} e^{-M_{IJ}(T-x_0)}\r]
\nn\ ,\eea
where the fit parameters are $\cZ_{IJ}$ and $M_{IJ}$, and $s_{IJ}=-1$ for $IJ=AP,PA$
and $+1$ otherwise. For all of our
quark masses, the fits are excellent. For each correlation function, we define the corresponding
pseudoscalar decay constant:
\bea
f_P^{IJ}&\equiv&\frac{m_1+m_2}{M_{IJ}^2}\sqrt{\cZ_{IJ}},\quad IJ=PP,PP{-}SS\nn\\
f_P^{IJ}&\equiv&Z_A\frac{\sqrt{\cZ_{IJ}}}{M_{IJ}},\quad IJ=AA,J^LJ^L\\
f_P^{IJ}&\equiv&\sqrt{Z_A\cZ_{IJ}\frac{m_1+m_2}{M_{IJ}^3}},\quad IJ=AP,PA\nn
\ ,\eea
where, in our conventions, $f_\pi=131\,\mev$.  The point of
considering these three definitions is that they differ in their
discretization errors. In fact, we expect these differences to be at
most equal to the mass-dependent discretization errors identified in
$Z_A(m_1,m_2)$. Indeed, $f_P^{PP}/f_P^{AA}$ is simply
$Z_A(m_1,m_2)/Z_A(0,0)$ to the extent that the fits to the
continuum-limit parametrizations of \eq{eq:2ptfndef} do not modify the
relative discretization errors in the amplitudes $\cZ_{IJ}$; that
$\nabla_0$ applied on $G_{A_0P}(t)$ yields a value of the pseudoscalar
mass which is the same as the one determined by the fits; that
finite-volume effects do not distort the various correlation functions
differently. And under these same conditions, all other ratios of
decay constants should be less than or equal to
$f_P^{PP}/f_P^{AA}$. Moreover, the correlation functions are affected
by topological zero modes differently. In particular, $PP-SS$ and
$J_0^LJ_0^L$ are free of zero modes. Differences at small quark masses
thus give a measure of zero-mode contamination.

\begin{figure}[!ht]
\vspace{-0.0cm}
\includegraphics*[width=7.5cm]{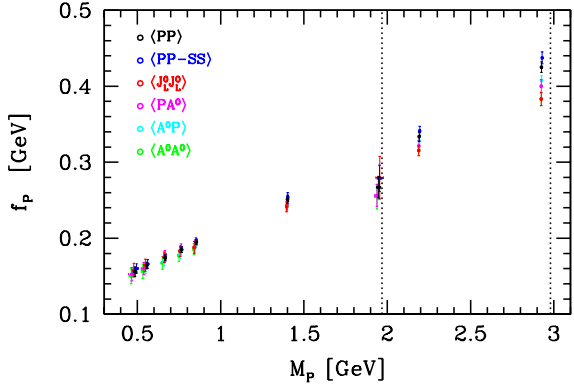}%
\vspace{-1.0cm}
\caption{$f_P$ vs.\ $M_P$ as obtained from various 2-point functions. For
clarity, only results for mesons composed of degenerate quarks are plotted, with the
exception of our fiducial $D_s$ meson. The vertical
lines correspond, from left to right, to $P=D_s$ and $P=\eta_c$. \label{fig:fPvsMP}}
\vspace{-0.6cm}
\end{figure}

In \fig{fig:fPvsMP} we plot $f_P^{IJ}$ versus $M_{IJ}$, for
all $IJ$. No statistically significant difference in the masses
obtained from the various correlation functions is observed. At low quark mass
this indicates that there is no evidence for zero-mode contamination in
our fits. At high quark mass, the agreement is reassuring but not
significant as discretization errors on the mass are determined by
the action, which is the same for all correlation functions. That is
not the case for the decay constant, where different correlation
functions will induce different discretization errors. Indeed, we find
that the difference increases as the quark mass is increased. For the
``$D_s$'', we find a variation of approximately 10\% and it is
15\% for the ``$\eta_c$'', which is entirely consistent with the results
obtained from $Z_A(m_1,m_2)$, as anticipated. While significant, these
variations do not indicate a major breakdown of the theory, even at
$am_1+am_2=1.5$. This is further confirmed by the fact that with a
charm quark such that the corresponding ``$D_s$'' meson has a mass $\sim
1.95\,\gev$ (i.e.\ $20\,\mev$ below experiment), our ``$\eta_c$'' has a
mass $\sim 2.93\,\gev$ which is only $50\,\mev$ (i.e.\ 2\%) below
experiment. Moreover, the preliminary value of $f_{D_s}$ that we obtain
is $267(15)(24)\,\mev$, which is in good agreement with
the most recent quenched calculation \cite{Juttner:2003ns} and experimental
measurement, $f_{D_s}=285(19)(40)\,\mev$ \cite{Heister:2002fp}.

\vspace{-0.2cm}

\subsection{Charm quark summary}

We have simulated the charm quark with overlap fermions on a lattice
with $a^{-1}\sim 2\,\gev$.  We have investigated the effects of
mass-dependent discretization errors on the axial Ward identity as
well as on the pseudoscalar decay constant. We find that these effects
are relatively small: 10\% for the ``$D_s$'' and 15\% for the
``$\eta_c$''.~\footnote{It should be noted, however, that these figures
come from ratios of quantities in which some discretization errors may cancel.} 
We further find a value of $f_{D_s}$ which is in good
agreement with the latest quenched and experimental results. Thus,
there is no sign of a major breakdown of the formalism even for bare
quarks masses $am_q=0.75$. Of course, these preliminary findings
require further investigation and, in particular, the apparent
smallness of discretization effects should be confirmed by simulations
on finer lattices.


\section{Weak matrix elements}

\vspace{-0.2cm}

\subsection{Bare matrix elements}


We have calculated the $K^0$-$\bar K^0$ mixing parameter $B_K$ following
the procedure detailed in \cite{Garron:2002dt,Garron:2003cb}. $B_K$
is defined as
\be
B_K(\mu)
=\frac{\langle
\bar K^0|O_{\Delta S=2}(\mu)|K^0
\rangle}{\frac{16}{3}M_K^2F_K^2}
\ee
with ($\gamma_\mu^{L,R}=\gamma_\mu(1\mp\gamma_5)$)
\be
O_{\Delta S=2}=\bar{s}\gamma_\mu^Ld\bar{s}\gamma_\mu^Ld
\ee

The bare matrix element $\langle\bar K^0|O_{\Delta S=2}|K^0\rangle$
is extracted from the $3$-point function
\be
C_{JOJ}(x_0,y_0) =
\sum_{\vec x,\vec y}\langle J^L_0(x) O_{\Delta S=2}^{bare}(0) J^L_0(y)\rangle
\ee
and the $2$-point function
\be
C_{JJ}(x_0) =  \sum_{\vec x}\langle J^L_0(x) \bar J^L_0(0)\rangle
\ee
which are both free of unphysical zero modes \cite{Giusti:2002sm}.

The large temporal extent of our lattice allows us to choose
ranges $10\le x_0/a\le 20$ and $44\le y_0/a\le 54$ for the kaon source
and sink, which are both safely in the asymptotic region $x_0\gg a$,
$y_0\ll T$ and for which backward contributions due to the
toroidal lattice geometry are negligible. As a consistency check, we
have verified that within these ranges the pseudoscalar meson mass
extracted from $C_{JOJ}(x_0,y_0)$ agrees very well with the ones
extracted from various $2$-point functions (cf.\ \sec{sect:lhs}).

We have also computed the bag parameters $B_{7/8}^{3/2}$ of the
electroweak penguin
operators in the $\Delta I=3/2$ channel of the $\Delta S=1$ effective
Hamiltonian. Following the convention of \cite{Lellouch:1998sg}, we define
\bea
B^{3/2}_{7}(\mu)=
\lim_{m\rightarrow 0}
\frac{\langle
\pi^+|O^{3/2}_{7}(\mu)|K^+
\rangle}{\frac{2}{3}\langle
\pi^+|P(\mu)|0\rangle\langle0|P(\mu)|K^+
\rangle}&&
\\
B^{3/2}_{8}(\mu)=
\lim_{m\rightarrow 0}
\frac{\langle
\pi^+|O^{3/2}_{8}(\mu)|K^+
\rangle}{2\langle
\pi^+|P(\mu)|0\rangle\langle0|P(\mu)|K^+
\rangle}&&
\eea
where $O^{3/2}_7$ is given by
\be
O^{3/2}_7=\bar{s}\gamma_\mu^Ld
(
\bar{u}\gamma_\mu^Ru-\bar{d}\gamma_\mu^Rd
)+
\bar{s}\gamma_\mu^Lu
\bar{u}\gamma_\mu^Rd
\ee
and $O^{3/2}_8$ is the corresponding color-mixed operator. We work in
the $SU(3)_V$ limit $m_u=m_d=m_s$, in which all eye contractions cancel.

In contrast to $B_K$, finding appropriate sources to cancel all
zero-mode contributions is not straightforward for $B_{7/8}^{3/2}$ due to
the $L$-$R$ structure of the operator. For this preliminary study,
we therefore choose to simply report results obtained with
pseudoscalar sources on the $3$-point function
\be
C_{7/8}(x_0,y_0) =
\sum_{\vec x,\vec y}\langle P(x) (O^{3/2}_{7/8})^{\text{bare}}(0) \bar
P(y)\rangle
\ee
and the $2$-point function
\be
C_{PP}(x_0) =  \sum_{\vec x}\langle P(x) \bar P(0)\rangle
\ee
and postpone a more careful investigation of this issue to a
future publication.\footnote{Pseudoscalar sources are not the optimal choice,
since they allow for $1/m^2$ divergences in the 3-point functions.}

We identify nice plateaus in $C_{7/8}(x_0,y_0)$ in the range $18\le
x_0/a\le 23$ and $41\le y_0/a\le 46$. In this region, the mass
extracted from the $3$-point function agrees very well with the mass
extracted from various $2$-point functions (cf.\ \sec{sect:lhs}).

\vspace{-0.2cm}

\subsection{Non-perturbative renormalization}

We perform all renormalizations non-perturbatively in the RI/MOM
scheme following \cite{Martinelli:1995ty}. Thus, we fix gluon
configurations to Landau gauge and numerically compute appropriate,
amputated forward quark Green functions with legs of momenta $p$.  The
renormalization of $B_K$ is detailed in \cite{Garron:2003cb}. In order to
renormalize $B_{7/8}^{3/2}$, we calculate the renormalization
constants of a full set of dimension $6$, $4$-flavor operators in the
basis suggested by \cite{Donini:1999sf}. In this basis, $O_7^{3/2}$
renormalizes as $Q^+_2$ and $O_8^{3/2}$ as $-2Q^+_3$. These $2$
operators mix among each other but chiral symmetry protects them from
mixing with other operators. The renormalization condition violates
chiral symmetry at low $p^2$, but we found, that the violation is
tiny.

The renormalization pattern for the $B$ parameters is given by
\be
\left(
\begin{array}{c}
B_{7}^{3/2}\\
B_{8}^{3/2}
\end{array}
\right)^{\text{RI}}\!\!\!\!\!\!(\mu)
=
\left(
\begin{array}{cc}
Z_{77} & Z_{78}\\
Z_{87} & Z_{88}
\end{array}
\right)^{\text{RI}}\!\!\!\!\!\!(\mu)
\left(
\begin{array}{c}
B_{7}^{3/2}\\
B_{8}^{3/2}
\end{array}
\right)^{\text{bare}}
\ee

In order to determine the $Z_{ij}^{\text{RI}}(\mu)$ we define
\be
\label{eq:Zri}
\cR_{ij}^{\text{RI}}(\mu)=\lim_{m\rightarrow 0}
\left.\frac{\Tr(P_S\Gamma_S(p^2,m))^2}{\Tr(P_i\Gamma_j(p^2,m))}
\right|_{p^2=\mu^2}
\ee
where $\Gamma_{i}(p^2,m)$ is the amputated Green function of the
operator $O_i^{3/2}$ and $P_i$ the corresponding projector
(cf. \cite{Donini:1999sf}). After performing an appropriate chiral
extrapolation, we must isolate the ``perturbative part'' of this ratio to
get the renormalization constant.  In order to do this,
we fit the matrix $\cR^{\text{RI}}(\mu)$ to the form
\bea
\cR_{ij}^{\text{RI}}(\mu)&=&\cdots+\frac{A_{ij}}
{\mu^4}+\frac{B_{ij}}{\mu^2}+Z_{ij}^{\text{RI}}(\mu)\nn\\
\label{eq:z78ope}
&&+C_{ij}\times(a\mu)^2+\cdots
\eea
where the running of $Z_{ij}^{\text{RI}}(\mu)$ is given by NLO
continuum perturbation theory. The $1/\mu^2$ and $1/\mu^4$ terms are
motivated by the OPE in $1/p^2$ of
\eq{eq:Zri}, while the term $\propto (a\mu)^2$ takes into account
leading order discretization effects.
\begin{figure}[!ht]
\includegraphics*[width=7.5cm]{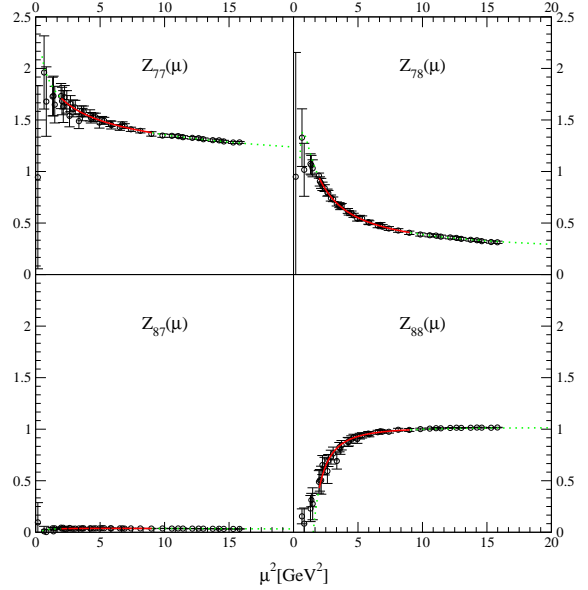}%
\vspace{-0.8cm}
\caption{\label{fig:z78}The coefficient matrix $\cR_{ij}^{\text{RI}}$
vs. $\mu$. One can see, that the data are very well described by
\eq{eq:z78ope}.}
\end{figure}

In \fig{fig:z78} the $\cR_{ij}^{\text{RI}}(\mu)$ are shown
together with fits of the form \eq{eq:z78ope} in a range
$\mu\in[\sqrt{2},3]\,\gev$. The fit describes the data very nicely even
far outside the fit range and is stable against addition of extra
terms and variations in the fit range. Some of the fit parameters have
numerically found to be consistent with $0$ and have been eliminated.

\vspace{-0.2cm}

\subsection{Physical results}

\begin{figure}[!ht]
\vspace{0.1cm}
\includegraphics*[width=7.5cm]{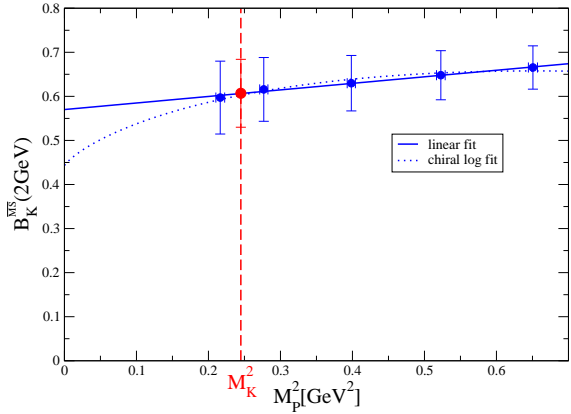}%
\vspace{-0.8cm}
\caption{\label{fig:bk}$B_K^{\overline{\text{MS}}\text{-NDR}}(2\text{GeV})$
vs.~the pseudoscalar meson mass. The continuous line gives the
result of a linear fit and the dotted line of a chiral log fit
of the form \eq{eq:chlog}.}
\end{figure}

In \fig{fig:bk} we show the renormalized value of $B_K$ in the
$\overline{\text{MS}}\text{-NDR}$ scheme at $2\,\gev$.
We have performed both a linear and a chiral log fit of the form
\be
B_K=B\l(1-6
\l(\frac{M}{4\pi F}\r)^2\ln
\frac{M}{\Lambda_B}\r)
\label{eq:chlog}
\ee
At the physical point, the result of both fits is virtually
indistinguishable and we obtain
$B_K^{\overline{\text{MS}}\text{-NDR}}(2\text{GeV})=0.61(7)$, where
the error is statistical only. This result is compatible both with our
earlier result from smaller lattices \cite{Garron:2003cb} and with a
recent determination by the MILC collaboration \cite{DeGrand:2003in}.

In \fig{fig:bk} we plot the renormalized value of $B^{3/2}_{7/8}$
in the $\overline{\text{MS}}\text{-NDR}$ scheme at $2\,\gev$ as a function
of $M_P^2$.
\begin{figure}[!ht]
\vspace{0.1cm}
\includegraphics*[width=7.5cm]{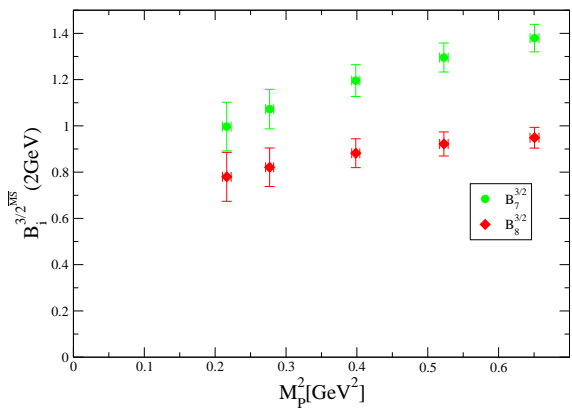}%
\vspace{-0.8cm}
\caption{\label{fig:b78}$\l(B^{3/2}_{7/8}\r)^{\overline{\text{MS}}\text{-NDR}}(2\text{GeV})$
vs. the pseudoscalar meson mass.}
\end{figure}
We postpone the study of this mass dependence to a later publication.

\bibliographystyle{my-elsevier}

\bibliography{proc}

\end{document}